\documentclass[prl,aps,superscriptaddress,footinbib,twocolumn]{revtex4-2}

\usepackage{graphicx}
\usepackage{dcolumn}
\usepackage{bm}
\usepackage[unicode=true,bookmarks=true,bookmarksnumbered=false,bookmarksopen=false,breaklinks=false,pdfborder={0 0 1},backref=false,colorlinks=true]
{hyperref}
\hypersetup{
	linkcolor=magenta, urlcolor=blue, citecolor=blue, pdfstartview={FitH}, hyperfootnotes=false, unicode=true}
\usepackage{url}
\usepackage{xcolor}

\usepackage[utf8]{inputenc}
\usepackage{amsmath}
\usepackage{txfonts}
\usepackage{amssymb} 
\usepackage{graphicx}
\usepackage{esint}
\usepackage{ulem}
\usepackage{multirow}
\usepackage{float}

\begin{document}

\title{Testing the unified bounds of the quantum speed limit
}

\affiliation{Zhejiang Key Laboratory of Micro-nano Quantum Chips and Quantum Control, School of Physics, and State Key Laboratory for Extreme Photonics and Instrumentation, Zhejiang University, Hangzhou 310027, China\\
$^2$Hangzhou Global Scientific and Technological Innovation Center, Hangzhou 311215, China\\
$^3$School of Mathematics, Physics and Statistics, Shanghai University of Engineering Science, Shanghai 201620, China\\
$^4$College of Optical Science and Engineering, Zhejiang University, Hangzhou 310027, China\\
$^5$Institute for Quantum Science and Engineering, Departments of Biological and Agricultural Engineering, Physics and Astronomy, Texas A$\&$M University, College Station, Texas 77843, USA
}

\author{Yaozu Wu$^{1}$}\thanks{These authors contributed equally to this work}
\author{Jiale Yuan$^{1}$}\thanks{These authors contributed equally to this work}\email{jialeyuan@zju.edu.cn}
\author{Chuanyu Zhang$^{1}$}
\author{Zitian Zhu$^{1}$}
\author{Jinfeng Deng$^{1}$}
\author{Xu Zhang$^{1}$}
\author{Pengfei Zhang$^{2}$}
\author{Qiujiang Guo$^{2}$}
\author{Zhen~Wang$^{1}$}
\author{Jiehui Huang$^3$}
\author{Chao Song$^{1}$}\email{chaosong@zju.edu.cn}
\author{Hekang Li$^{2}$}\email{hkli@zju.edu.cn}
\author{Da-Wei Wang$^{1, 4}$}
\author{H.Wang$^{1, 2}$}
\author{Girish S. Agarwal$^{5}$}

\date{\today}

\begin{abstract} 
Quantum speed limits (QSLs) impose fundamental constraints on the evolution speed of quantum systems.
Traditionally, the Mandelstam-Tamm (MT) and Margolus-Levitin (ML) bounds have been widely employed, relying on the standard deviation and mean of energy distribution to define the QSLs.
However, these universal bounds only offer loose restrictions on the quantum evolution.
Here we introduce the generalized ML bounds, which prove to be more stringent in constraining dynamic evolution, by utilizing moments of energy spectra of arbitrary orders, even noninteger orders.
To validate our findings, we conduct experiments in a superconducting circuit, where we have the capability to prepare a wide range of quantum photonic states and rigorously test these bounds by measuring the evolution of the system and its photon statistics using quantum state tomography.
While, in general, the MT bound is effective for short-time evolution, we identify specific parameter regimes where either the MT or the generalized ML bounds suffice to constrain the entire evolution.
Our findings not only establish new criteria for estimating QSLs but also substantially enhance our comprehension of the dynamic evolution of quantum systems.
\end{abstract}

\maketitle
The quantum speed limits (QSLs) set fundamental bounds on how fast a quantum state can evolve,
typically determined by the energy spectra of the quantum states. 
They provide essential guidance in improving the performance of quantum devices in quantum computing and gate engineering \cite{lloyd2000,lloyd2002,svozil2005,santos2015,Aifer_2022}, quantum optimal control \cite{carlini2006,caneva2009,bason2012,frank2016}, and quantum information processing \cite{murphy2010}.
Two famous bounds, the Mandelstam-Tamm (MT) bound \cite{mt1945,PhysRevLett.65.1697} and the Margolus-Levitin (ML) bound \cite{margolus1998maximum,kosi2006,levitin2009,Giovannetti2003,anaproof}, which are determined by the standard deviation and mean of the energy spectra, have been experimentally tested in an infinite-level system \cite{doi:10.1126/sciadv.abj9119}. These two bounds limit the dynamical evolution in different parameter regimes and at different evolution times \cite{doi:10.1126/sciadv.abj9119}. By considering the time-reversal dynamics, a dual ML bound was introduced for systems with finite energy spectra \cite{PhysRevLett.129.140403}. To choose a proper bound to estimate the speed limit, three parameter regimes of the variance and mean energy, restricted by each of these three bounds, have been found. {However, in the ML and dual ML regimes, the QSLs are still more tightly bounded by the MT bound before a crossover time. Moreover, these bounds are in general not tight, i.e., the bounds do not touch the evolution curve of the wavefunction overlap.}
Since the dynamic evolution is governed by the energy spectra and only the first and second order moments have been used in those bounds, it is natural to ask whether other moments of the energy spectra can put tighter bounds to the QSLs \cite{luo2005,gml2006}. 

Here we use arbitrary order moments (the so-called $L^p$-norms) to obtain a series of generalized ML (GML) bounds that can tightly constrain the quantum evolution. With this generalization, the previously found ML, dual ML, and MT regimes are tightly constrained by the corresponding (generalized) bounds, such that we only need to use one of them to estimate the QSLs. To experimentally validate our findings, we use a superconducting circuit \cite{krantz2019quantum, you2011atomic} to test the tightness of the bounds. 
Quantum states with different photon statistics of a free resonator are prepared and their evolution is measured by the quantum state tomography (QST), such that we obtain not only the overlap between the final and initial states, but also the values of the energy moments and the corresponding bounds. We test the universality of the bounds for classical and nonclassical states, including the two-, three- and infinite-level systems. In particular, we focus on the important photonic states including coherent states, squeezed states and superposition of Fock states. These states belonging to different classes can have the same mean energy, and yet the moments can be very different, which allows us to examine the universality of the generalized QSL bounds. This study substantially extends our ability in estimating QSLs and provides new guidance in quantum information processing.

\begin{figure*}[t]
\centering
\includegraphics[width=1.0\textwidth]{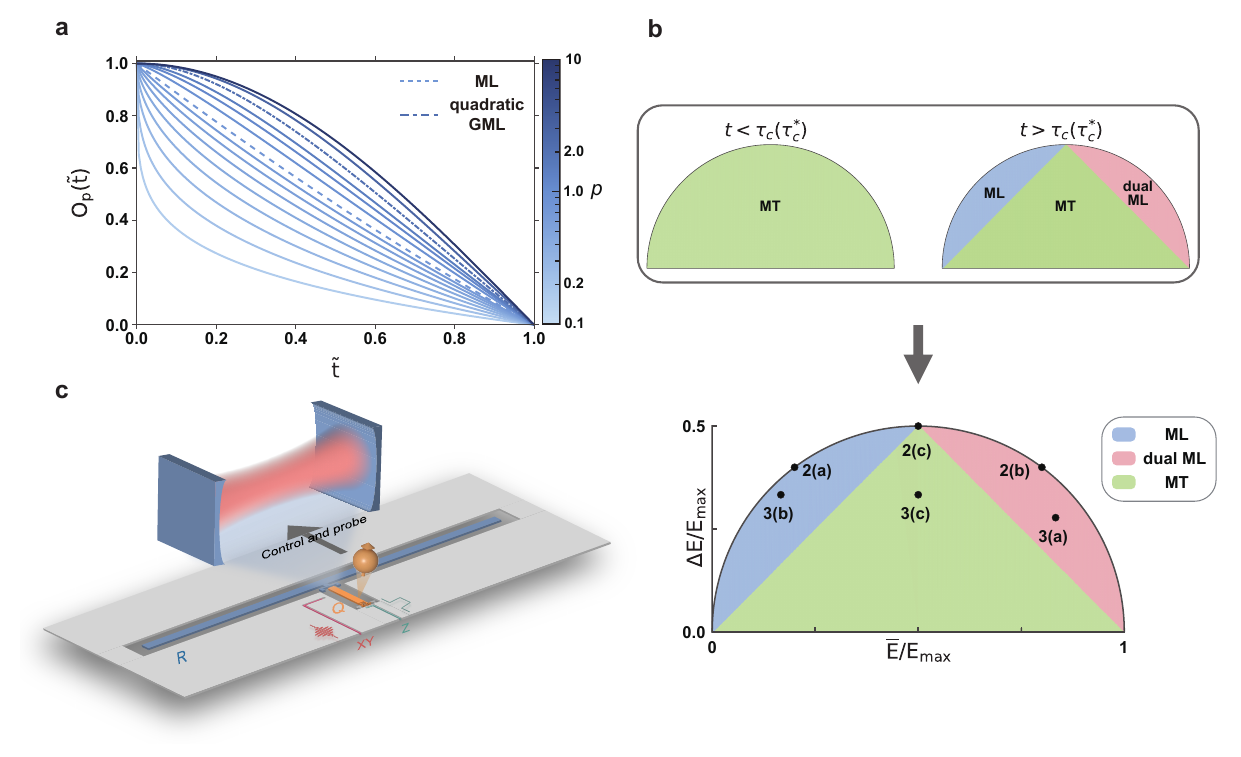}
\caption{\label {Fig1} {\bf Generalized quantum speed limits and their experimental tests.}
(a) The family of the GML bounds. We plot the GML bounds $O_p(\tilde{t})$'s for $0.1\leq p\leq10$.
The ML bound with $p=1$ is shown by the dashed line and the quadratic GML bound with $p=2$ is shown by the dash-dotted line.
{(b) The classification of the parameter regimes according to the known [MT, ML, dual ML] (upper) and our generalized (lower) bounds.
In the known framework, there exits a crossover time $\tau_c$ ($\tau_c^{*}$), before which the evolution are bounded by the MT bound.
After $\tau_c$ ($\tau_c^{*}$), the QSLs are bounded by the MT (green), ML (blue) and dual ML (red) bounds in the corresponding parameter regimes. With the generalized bounds, $\tau_c$ ($\tau_c^{*}$) approaches to zero and a single parameter diagram suffices to determine the tightest bound.
The energy moments of the states investigated in Fig.\ref{Fig2} and Fig.\ref{Fig3} are marked in the diagram.
(c) A schematic of the experimental setup used to test the unified bounds of QSLs. The experiment is performed on a superconducting circuit device composed of a coplanar waveguide resonator (R) and a transmon qubit (Q). The qubit, which can be flexibly manipulated by two control lines, i.e., an XY line for microwave control and a Z line for frequency tunability, is used to realize arbitrary state preparation and measurement of the resonator. The whole system resembles that of a cavity quantum electrodynamics device, with the resonator playing a role of cavity, and the qubit an artificial atom to control and probe the quantum state stored in the cavity.}
}
\label{fig_p_bound}
\end{figure*}

\noindent\textbf{The Unified Bounds of QSLs} 
\newline\noindent
The MT bound constrains the QSLs with the standard deviation of the energy,
\begin{equation}
\begin{aligned}
|\langle \psi(0)|\psi(t)\rangle| \geq  \cos\left(\frac{\Delta E t}{\hbar}\right),
\end{aligned}
\end{equation}
where $\Delta E=(\langle \hat{H}^2\rangle-\langle \hat{H}\rangle^2)^{1/2}$. The bound is valid when $0\leq t\leq \tau_{\text{MT}}$ with $\tau_{\text{MT}}\equiv \pi\hbar/(2\Delta E)$ being the MT minimal orthogonality time. We define $t_{\bot}$ as the shortest time required for a state evolving to one of its orthogonal states. Constrained by the MT bound, $t_{\bot}\geq \tau_{\text{MT}}$.
On the other hand, the ML bound constrains the evolution with the mean energy of quantum states, $t_{\bot}\geq \pi\hbar/(2 \Bar{E})$ \cite{margolus1998maximum}, where the mean energy $\Bar{E}=\langle H-\epsilon_0\rangle$ and $\epsilon_0$ is the ground state energy. 

\begin{figure*}[th]
	\includegraphics[width=1.0\textwidth]{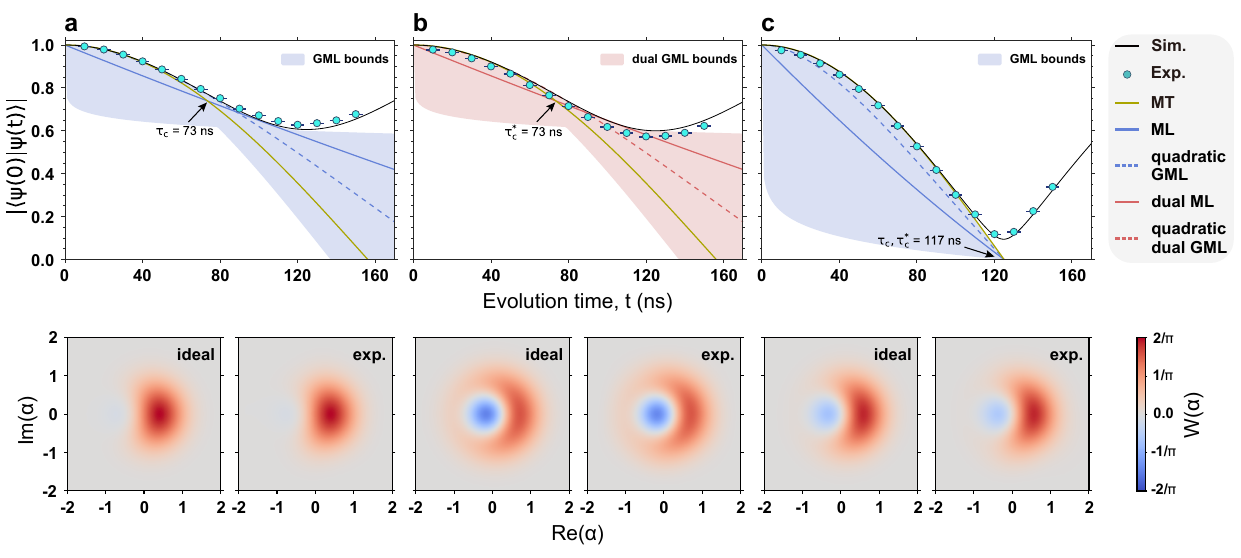}
	\caption{\label {Fig2} {\bf QSLs of two-level systems.} 
 We prepare the initial states $(2|0\rangle+|1\rangle)/\sqrt{5}$ (a), $(|0\rangle+2|1\rangle)/\sqrt{5}$ (b), and $(|0\rangle+|1\rangle)/\sqrt{2}$ (c), which correspond to the points in the ML regime, dual ML regime, and at the critical point as marked in Fig.\ref{fig_p_bound}(b), respectively. 
 The fidelities of the three initial states, defined as $F=\langle \psi|\rho|\psi \rangle$ with uncertainty estimated from the fluctuation of measured qubit population, are $F = 0.9978\pm 0.0007, 0.9796\pm 0.0014$, and $0.9787\pm 0.0010$, respectively.
 Colored regions are delineated by the GML (blue) and dual GML (red) bounds with $0.1\leq p\leq10$.
 In the three cases, the tightest constraint on the QSLs are all given by the GML or dual GML bounds.
 The measured evolution data is marked by the light blue dots, and the simulated evolution data is shown by the black lines.
 The theoretical crossover time between the MT and (dual) ML bounds are illustrated by ($\tau_c^{*}$) $\tau_c$.
 In the lower panel, we display both the theoretical and the measured Wigner functions for the initial states.
 The measured Wigner functions have been adjusted to compensate the phase delay introduced by the XY control line of the ancilla qubit.
 }
\end{figure*}

The MT and ML bounds only involve with the variance and mean of the energy spectra. For Gaussian spectra, these 
are sufficient to describe the statistics of the energy and thus the quantum evolution of the system. However, in most cases, other moments of the energy spectra are needed. It was found that the evolution is bounded by a family of $p$th order moment of the energies, $E_p = \langle (\hat{H}-\epsilon_0)^p\rangle^{1/p}$, such that $t_{\bot}\geq \tau_p\equiv \pi\hbar/(2^{1/p}E_p)$, where $p$ can be any positive number \cite{gml2006}, not necessarily an integer.
Based on the orthogonality time of the GML bounds, we obtain the constraints that limit the overlap between the final and initial states during the evolution,
\begin{equation}
\begin{aligned}
|\langle \psi(0)|\psi(t)\rangle| \geq O_{p}\left(\frac{t}{\tau_p}\right).
\end{aligned}
\label{GML}
\end{equation}
The functions $O_p$'s defined in the time domain $0\leq t/\tau_p\leq 1$ decrease from $1$ to $0$ (see Supplementary Information (SI) and Fig.\ref{fig_p_bound}(a)).
For $p=1$, Eq.~(\ref{GML}) reduces to the conventional ML bound. 

The dual GML bound, valid for systems with a \textit{maximum} energy $\epsilon_{\text{max}}$, is obtained from the GML bound by replacing the energy moments with their dual values,
\begin{equation}
\begin{aligned}
|\langle \psi(0)|\psi(t)\rangle| \geq O_{p}\left(\frac{t}{\tau_p^*}\right),
\end{aligned}
\end{equation}
where $E_p^*\equiv\langle (\epsilon_{\text{max}}-\hat{H})^p\rangle^{1/p}$ is the dual $p$th order moment of the energy, and $\tau_{p}^* \equiv \pi\hbar/(2^{1/p}E_p^*)$ is the dual GML minimal orthogonality time.  The MT, GML,  and dual GML bounds put a unified bound on the QSLs.

To determine the tightest bound, we compare their orthogonality times and obtain a diagram of the mean and variance of the energy spectra (see Fig.\ref{fig_p_bound}(b)).
Conventionally, the parameter space of $\bar{E}$ and $\Delta E$ can be classified in three regimes \cite{PhysRevLett.129.140403}.
The ML and dual ML regimes are determined by the relations $\tau_{\text{ML}}>\tau_{\text{MT}}$ and $\tau_{\text{ML}}^{*}>\tau_{\text{MT}}$, such that $\bar{E}<\Delta E$ and $\bar{E}^*<\Delta E$, respectively. However, in these two regimes, the evolution is initially limited by the MT bound and later by the (dual) ML bound after a crossover time ($\tau_c^*$) $\tau_c$, i.e., the time when the (dual) ML and MT bounds cross. Therefore, we still need to consider both the MT and (dual) ML bounds in the (dual) ML regime. On the other hand, in the MT regime, we only need to consider the MT bound which puts the tightest constraint.

By considering the $p$th order moment, however, the (dual) ML regime can be tightly restricted by the (dual) GML bounds, while the MT bound can be discarded. In the MT regime, the MT bound still provides the tightest constraint.
Therefore, after the generalization of the (dual) ML bound, we can use only one type of bounds to estimate the QSLs in each parameter regime. The crossover times $\tau_c$ and $\tau_c^*$ become irrelevant for our estimation. In the following, we experimentally test these parameter regimes for quantum systems with different number of energy levels as well as different energy statistics.

The experiment is performed on a superconducting circuit, where a tunable transmon qubit is directly coupled to a fixed frequency resonator \cite{Wang2020}. 
The transmon qubit serves as an ancilla qubit for both preparing and detecting the desired state within the resonator.
Utilising the ancilla qubit, any nonclassical superposition of Fock states $|\psi(0)\rangle=\sum_{n}c_n|n\rangle$ can be prepared in the resonator (see SI) \cite{2009Synthesizing}.
After obtaining the initial state, we detune the ancilla qubit from the interaction point, allowing the resonator to evolve freely under the Hamiltonian $H=\hbar\nu a^\dagger a$.
Here, the effective frequency is $\nu \approx 2\pi\times4$ MHz (see SI).
At different times of the evolution, we perform QST to the resonator to acquire the density matrices and the corresponding Wigner functions $W_t(x,p)$ \cite{PhysRev.40.749}. 
The overlap is obtained by $|\langle\psi(0)|\psi(t)\rangle| = [\text{Tr}(\rho_0\cdot\rho_{t})]^{1/2}=[\int_{-\infty}^{\infty} W_0(x,p)W_t(x,p)dx^2dp^2]^{1/2}$.

\noindent\textbf{Two-level System} 
\newline\noindent Two-level systems are important for quantum computing \cite{arute9811,gong7812,google2023suppressing,takeda2022quantum}. 
{The energy moments of two-level states belong to either the ML or the dual ML regime, where the unified GML or dual GML bounds solely determine the QSLs during the whole evolution.}
We present three representative examples in Fig.\ref{Fig2} referring to states in the ML dominated regime, the dual ML dominated 
regime, and at the critical point. The energy moments of these states are marked in Fig.\ref{fig_p_bound}(b).
In general, for a state $|\psi\rangle=\cos\theta|0\rangle+e^{\text{i}\phi}\sin\theta|1\rangle$, the standard deviation, the mean, and the dual mean of the energy are $\Delta E=\hbar\nu\sin\theta\cos\theta$, $\Bar{E}=\hbar\nu\sin^2\theta$ and $\Bar{E}^*=\hbar\nu\cos^2\theta $. The minimal orthogonality times evaluated by the MT, ML, and dual ML bounds are $\tau_{\text{MT}}=\pi/(2\nu\sin\theta\cos\theta)$, $\tau_{\text{ML}}=\pi/(2\nu\sin^2\theta)$, and $\tau_{\text{ML}}^*=\pi/(2\nu\cos^2\theta)$, respectively. For $\theta>\pi/4$, we have $\tau_{\text{ML}}<\tau_{\text{MT}}<\tau_{\text{ML}}^*$, otherwise the opposite. At the critical point where $\tau_{\text{ML}}=\tau_{\text{MT}}=\tau_{\text{ML}}^*$, we obtain $\theta=\pi/4$. 

\begin{figure}[htp]
\includegraphics[width=0.5\textwidth]{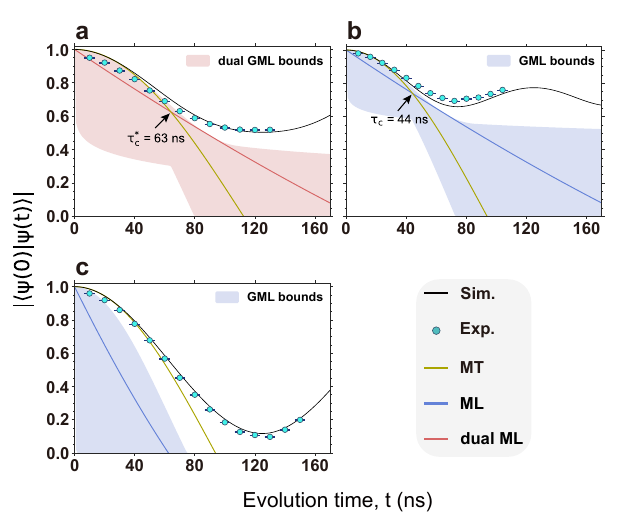}
\caption{\label {Fig3} {\bf QSLs of three-level systems.} 
We prepare the initial states $(\sqrt{7}|0\rangle+\sqrt{40}|1\rangle+\sqrt{115}|2\rangle)/\sqrt{162}$ (a), $(\sqrt{7}|0\rangle+|1\rangle+|2\rangle)/3$ (b), and $(\sqrt{2}|0\rangle+\sqrt{5}|1\rangle+\sqrt{2}|2\rangle)/3$ (c), which correspond to the points in dual ML, ML, and MT regimes in Fig.\ref{fig_p_bound} (b), respectively. The corresponding Wigner functions and fidelities of the three initial states are provided in SI. The evolutions are mainly restricted by the dual GML, GML, and MT bounds in the three cases. The measured evolution data is marked by the light blue dots, and the simulated evolution data is shown by the black lines.
 }
\end{figure}

\begin{figure*}[th]
\includegraphics[width=1.0\textwidth]{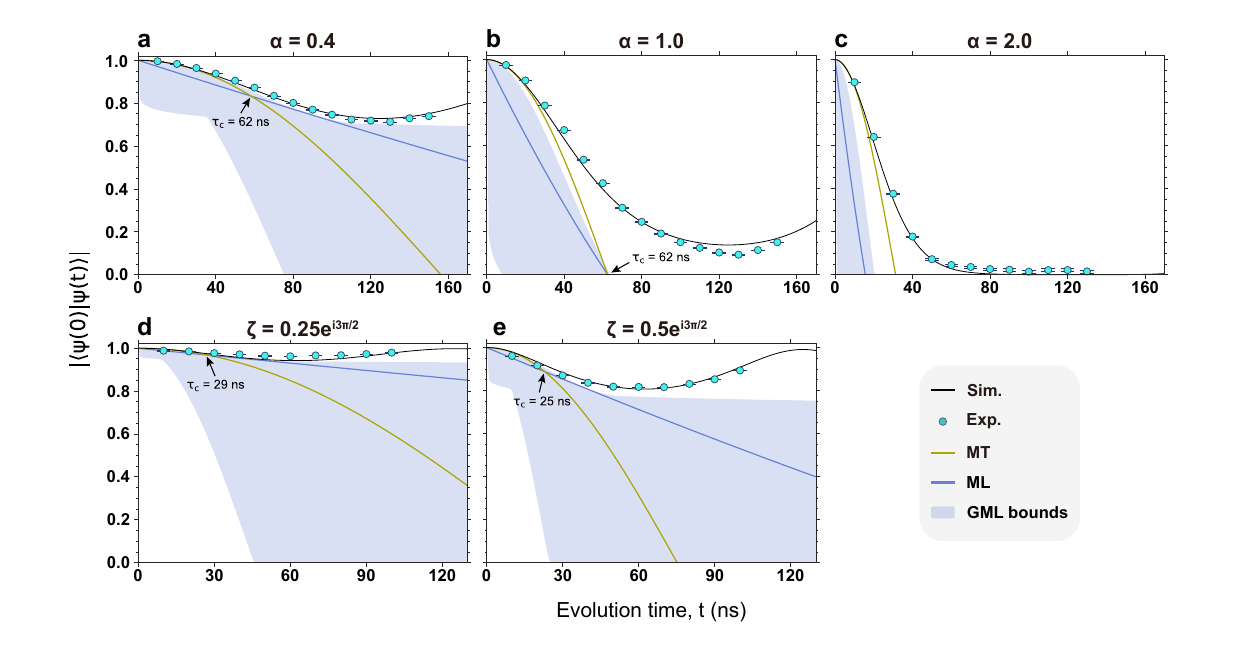}
\caption{\label {Fig4} {\bf QSLs of coherent states and squeezed states.}
(a-c) QSLs of coherent states with $\alpha=0.4, 1.0$, and 2.0, whose evolutions are observed to be in the ML regime, critical point, and MT regime, respectively. The QSLs are mainly governed by the GML (a) and MT (b and c) bounds, respectively.
(d) and (e) QSLs of squeezed states with $\zeta=0.25e^{i3\pi/2}$ and $0.5e^{i3\pi/2}$.
The QSLs are governed by the GML bounds.
The corresponding Wigner functions and fidelities of the three initial states are provided in SI. The measured evolution data is marked by the light blue dots, and the simulated evolution data is shown by the black lines.
}
\end{figure*}

In previous studies, we need to obtain the crossover time $\tau_c$ and $\tau_c^*$ to determine which bound to use for different evolution times (see Fig.\ref{Fig2}(a))  \cite{doi:10.1126/sciadv.abj9119,PhysRevLett.129.140403}. 
With the generalization of the ML bounds, the whole evolution of two-level systems is governed by the GML bounds with different $p$'s, leaving the MT bound unnecessary in evaluating the QSLs. The new bounds are more tight. We can always find a bound that touches the overlap function before the time of the minimum overlap. This is consistent with the theoretical analysis (see SI), which can be generalized to arbitrary two-level states. 
For the GML bounds with $p<1$, their $O_p$ curves are concave up and limit the evolution at long time. Their effective regimes are characterized by the minimal orthogonality time $\tau_p=\pi/(2^{1/p}\nu \sin^{2/p}\theta)$ \cite{gml2006}. For $p=0.1$ in Fig.~\ref{Fig2}(a), we obtain $\tau_{0.1}/\tau_{\text{ML}}=3.8\times10^3$, indicating that the GML bound can restrict the evolution for a much longer time than the 
{conventional ML} bound. 
On the other hand, for GML bounds with $p>1$, the $O_p$ curves are concave down and limit the short time evolution.

By inverting the populations of the two energy levels, we prepare the initial state in the dual ML regime (see Fig.\ref{Fig2}(b)).
Unnecessary to invoke $\tau_c^*$, the dual GML bound tightly limits the evolution.
Similar to the ML regime, the dual GML bounds with $p<1$ and $p>1$ limit the long and short time evolution, respectively. 
When the populations in the two energy levels are equal, $c_0=c_1$, we reach to the critical point where all bounds have the same orthogonality time \cite{gml2006} (see Fig.\ref{Fig2}(c)). The theoretical results show that the MT, GML, and dual GML bounds are all tight in the limit $p\rightarrow\infty$. At this critical point, the GML and the dual GML bounds are the same.

An important question is on the range of $p$ for a confident estimation of the QSLs from the GML bounds. For $p<1$, the GML bounds are complement to the MT bound, providing tighter constraint for long time evolution. However, for $p>1$, MT bound can put a tight constraint for small time, such that GML bounds with $p\gg 1$ are in general unnecessary (see SI). 
A {generally} useful one is the quadratic GML bound with $p=2$. As shown in Fig.\ref{Fig2}(a) and (b), the (dual) quadratic GML bounds are tighter than the MT and (dual) ML bounds at the crossover time ($\tau_c^{*}$) $\tau_c$. Moreover, the second order energy moment can be obtained from the energy moment by $E_2 =\sqrt{\Delta E^2+E^2}$, which can be measured in simpler ways than QST \cite{doi:10.1126/sciadv.abj9119}.

\noindent\textbf{Multi-level System} 
\newline\noindent
For two-level systems, we can discard the MT bound and only use the GML bounds to estimate the QSLs. This is not necessarily true for multi-level systems with more complicated energy statistics. In the following, we use three-level systems to investigate the QSLs in all three dynamic regimes, i.e., the dual ML, ML, and the MT regimes in Fig.\ref{fig_p_bound}(b). We find that in the dual ML and ML regimes the QSLs are still governed by the unified GML bounds for the whole evolution, while in the MT regime by the MT bound. The MT bound might be tighter than the GML bounds in the (dual) ML regime for short time, but in general such a difference can be neglected.
In Fig.\ref{Fig3}(a) and (b), we prepare the initial states in the dual ML and ML regimes, respectively. The QSLs are solely determined by the dual GML and GML bounds, while the MT bound can be discarded (see SI).
In Fig.\ref{Fig3}(c), we prepare the initial state in the MT regime, where the MT bound provides the tight constraint during the evolution. 

\noindent\textbf{Infinite-level System} 
\newline\noindent 
We next test the bounds in infinite-level systems. For systems without a maximum energy, we only consider the MT and GML bounds. {We find that in the ML regime, the QSLs are also governed by the GML bounds and the MT bound can be neglected.}
We first investigate the QSLs of the coherent states $|\alpha\rangle$  \cite{PhysRevLett.107.083601,PhysRevLett.111.033603,PhysRevLett.96.060504}.
The amplitude of $\alpha$ determines the parameter regimes.
The coherent state with small amplitude belongs to the ML regime as it is like a superposition of few Fock states and otherwise the MT regime because of the Gaussian nature of state (see SI).  
The critical point is $|\alpha|=1$, determined by $\tau_{\text{ML}}=\tau_{\text{MT}}$.

To experimentally prepare the coherent states, we use the XY control line of the ancilla qubit to displace the resonator with proper pulse amplitudes. 
In Fig.\ref{Fig4}(a-c), we show the evolution in the ML regime, at the critical point, and in the MT regime, respectively.
For small $|\alpha|$ in Fig.\ref{Fig4}(a), the QSLs are governed by the GML bounds. 
When $\alpha=1$, the GML and MT bounds provide similar constraints (see Fig.\ref{Fig4}(b)). 
For a large $|\alpha|$ the MT bound constrains the QSLs (see Fig.\ref{Fig4}(c)).

We next test the bounds for the squeezed states, which are important for quantum metrology \cite{sq1,sq2,sq4,sq6,sq7,PhysRevApplied.15.044030,Li:22}. Theoretical calculations show that the squeezed states $|\zeta\rangle$ are always in the ML regime for the free evolution Hamiltonian (see SI).
In our experiments, the squeezed states are realized by preparing superposition of Fock states with a photon number cutoff $N=6$ (see SI).
We show two examples in Fig.\ref{Fig4}(d) and (e) with $\zeta=0.25e^{i3\pi/2}$ and $0.5e^{i3\pi/2}$.
The QSLs are governed by the GML bounds, which is also true for other values of $\zeta$.
The GML bounds are tighter for smaller $|\zeta|$ and are naturally applicable as the generated squeezed state is in finite dimensional Hilbert space.
We also implement experiments for the squeezed coherent states (see SI).

\noindent\textbf{Conclusions} 
\newline\noindent {We generalize the unified bounds of QSLs by using arbitrary order energy moments.
We show that three parameter regimes determined by the mean and variance of the energy spectra can be used to determine which (generalized) bounds can provide the tightest constraint on the QSLs. 
In each regime, the QSLs are mainly governed by the corresponding bounds during the whole evolution. This provides simple and powerful guidance on estimating the QSLs for states with various energy spectra.
We experimentally test our theoretical findings with the bosonic states of a resonator in a superconducting circuit.
The results demonstrate the validity of the parameter regimes we have classified. 
This work substantially extends our ability in estimating the speed of quantum evolution and provides necessary guidance in designing quantum gate and devices.
This work can be further extended to quantum open systems by using the norms of the nonunitary operators \cite{campo2013,shanahan2018,deffner2013,taddei2013}.

\vspace{.5cm}
	\noindent\textbf{Acknowledgements} 
The device was fabricated at the Micro-Nano Fabrication Center of Zhejiang University.
We acknowledge the support of the National Natural Science Foundation of China (Grants No. 12174342, 12274368, 12274367, 92065204, and U20A2076), 
and the Zhejiang Province Key Research and Development Program (Grant No. 2020C01019).
G.S.A thanks the Air Force Office of Scientific Research (Award No. FA-9550-20-1-0366) and the Robert A. Welch Foundation (Grant No. A-1943-20210327) for supporting this work.
	
\vspace{.3cm}
\noindent\textbf{Author contributions} G.S.A. proposed the project. Y.W and C.Z. carried out the experiments and analyzed the experimental data under the supervision of C.S. and H.W.. X.Z. designed the device and H.L. fabricated the device. J.Y., J.H., D.W.W and G.S.A. conducted the theoretical analysis. All authors contributed to the experimental set-up, analysis of data, discussions of the results, and writing of the manuscript.

\bibliography{QSL}

\end{document}



\title{Supplementary Information for ``Testing the unified bounds of the quantum speed limit"\\ }

\affiliation{Zhejiang Key Laboratory of Micro-nano Quantum Chips and Quantum Control
, School of Physics, and State Key Laboratory for Extreme Photonics and Instrumentation, Zhejiang University, Hangzhou 310027, China\\
$^2$Hangzhou Global Scientific and Technological Innovation Center, Hangzhou 311215, China\\
$^3$School of Mathematics, Physics and Statistics, Shanghai University of Engineering Science, Shanghai 201620, China\\
$^4$College of Optical Science and Engineering, Zhejiang University, Hangzhou 310027, China\\
$^5$Institute for Quantum Science and Engineering, Departments of Biological and Agricultural Engineering, Physics and Astronomy, Texas A$\&$M University, College Station, Texas 77843, USA
}

\author{Yaozu Wu$^{1}$}\thanks{These authors contributed equally to this work}
\author{Jiale Yuan$^{1}$}\thanks{These authors contributed equally to this work}\email{jialeyuan@zju.edu.cn}
\author{Chuanyu Zhang$^{1}$}
\author{Zitian Zhu$^{1}$}
\author{Jinfeng Deng$^{1}$}
\author{Xu Zhang$^{1}$}
\author{Pengfei Zhang$^{2}$}
\author{Qiujiang Guo$^{2}$}
\author{Zhen~Wang$^{1}$}
\author{Jiehui Huang$^3$}
\author{Chao Song$^{1}$}\email{chaosong@zju.edu.cn}
\author{Hekang Li$^{2}$}\email{hkli@zju.edu.cn}
\author{Da-Wei Wang$^{1, 4}$}
\author{H.Wang$^{1, 2}$}
\author{Girish S. Agarwal$^{5}$}
 

\begin{abstract} 
\end{abstract}

\maketitle
\section{The extended generalized Margolus-Levitin bound}

Previous works \cite{gml2006,kosi2006} have shown the orthogonality times of the family of generalized Margolus-Levitin (GML) bounds, 
$\tau_{p}=\pi\hbar/(2^{1/p}E_p)$, in terms of the $L^p$-norm of the lower-bounded Hamiltonian $E_p = \langle (\hat{H}-\epsilon_0)^p\rangle^{1/p}$  with $\epsilon_0$ to be the ground state energy. 
Here, $p$ takes any positive number.
We extend the GML bounds applicable for an arbitrary overlap between the initial and the evolved states 
$|\langle \psi(0)|\psi(t)\rangle| \geq O_{p}\left(t/\tau_p\right)$.
To find the exact expression of $O_p$, it is sufficient to focus on two-level systems (as we will prove). Without loss of generality, we assume the ground state energy to be zero such that the two-level Hamiltonian is $H=\epsilon_1|1\rangle\langle 1|$. For an initial state $|\psi(0)\rangle = \sqrt{1-\xi^2}|0\rangle + \xi |1\rangle$ with $E_p = \xi^{2/p}\epsilon_1$, the evolving overlap is
\begin{equation}
|\langle \psi(0)|\psi(t)\rangle| = \left(1-2\xi^2(1-\xi^2)(1-\cos \epsilon_1t)\right)^{1/2}.
\end{equation}
By substituting $E_p$ in replacement of $\epsilon_1$ and minimizing the above expression over $z\equiv\xi^2$, we obtain the GML bound
\begin{equation}
O_p(\tilde{t}) = \mathop{\text{min}}\limits_{z\in[0,1]}\left( 1-2z(1-z)(1-\cos[(2z)^{-1/p}\tilde{t} \pi]) \right)^{1/2},
\label{eq:gmlbounds}
\end{equation}
where $\tilde{t}=t/\tau_p=2^{1/p}E_pt/(\pi\hbar)$ is the dimensionless reduced time. This is the main result of this section. The auxiliary $z$ is determined by the following equation
\begin{equation}
(1-2z)(1-\cos(\pi \tilde{t}(2z)^{-1/p}))-(1-z)\pi \tilde{t}\frac{1}{p}(2z)^{-\frac{1}{p}}\sin(\pi \tilde{t}(2z)^{-1/p})=0.
\label{eq:constr}
\end{equation}
The zero point for the above equation is not unique and to obtain the global minimum we take the one closest to (and less than) $0.5$.

In the remaining part of this section, we will prove that the bound is indeed taken for two-level states, which completes our proof for the GML bound.
Inspired by the work \cite{anaproof}, we study an equivalent problem as seeking the minimal time $T_{\delta}(E_p)$ required for an evolving state to attain an overlap $\sqrt{\delta}$ with energy $E_p$. We normalize it by the orthogonality time $T_{\delta}(E_p)/\tau_p$ and we have
\begin{equation}
\gamma_p(\delta) \equiv T_{\delta}(E_p) /\tau_p= \mathop{\text{inf}}\limits_{(H,\rho,\tau)}\frac{2^{1/p}}{\pi\hbar}\tau\langle (H-\epsilon_0)^p\rangle^{1/p},
\label{dyndis}
\end{equation}
where the infimum is over triples that the Hamiltonian $H$ evolves the initial state $\rho$ to a final state with overlap $\sqrt{\delta}$, at time $\tau$.
We intend to prove that the minimum value (if exists) is taken for two-level systems.
Without loss of generality, we assume $\epsilon_0=0$ and we focus on the unproved cases for $p\neq 1$ and $\delta>0$ (see \cite{anaproof} for $p=1$ and \cite{gml2006,kosi2006} for $\delta=0$). To simplify Eq.~(\ref{dyndis}), we  normalize the time  by defining $H_2=\tau H$ that evolves the initial state to the same final state at time $\tau=1$ through the same path. Then we mod the eigenvalues of $H_2$ by $2\pi$ to construct $H_3$ that evolves the state to the same final state but with smaller $L^p-$norm of the Hamiltonian. This means that $\gamma_p$ must be taken for systems whose eigenenergies are strictly less than $2\pi$.
As a result, we have
\begin{equation}
\gamma_p(\delta) = \mathop{\text{inf}}\limits_{(H,\rho)}C_p\langle (H)^p\rangle^{1/p},
\label{eq:timeplus}
\end{equation}
where $C_p=2^{1/p}/(\pi\hbar)$ is a constant multiplier and $\text{Tr}(\rho e^{-iH}\rho e^{iH})=\delta$ is the dynamical constraint.

Define functions $f(\boldsymbol{\rho},\boldsymbol{\epsilon})$, $g(\boldsymbol{\rho},\boldsymbol{\epsilon})$, $h(\boldsymbol{\rho},\boldsymbol{\epsilon})$ on $M=[0,1]^{2n}\times[0,2\pi]^n$ as,
\begin{equation}
\begin{aligned}
&f(\boldsymbol{\rho},\boldsymbol{\epsilon})=\sum_{i=1}^{n}\rho_i\epsilon_i^p,\\
&g(\boldsymbol{\rho},\boldsymbol{\epsilon})=\left(\sum_{i=1}^{n}\rho_i\cos\epsilon_i\right)^2+\left(\sum_{i=1}^{n}\rho_i\sin\epsilon_i\right)^2,\\
&h(\boldsymbol{\rho},\boldsymbol{\epsilon})=\sum_{i=1}^{n}\rho_i.
\end{aligned}
\end{equation}
The subset $M'\subset M$ defined by the constraints $g(\boldsymbol{\rho},\boldsymbol{\epsilon})=\delta$ and $h(\boldsymbol{\rho},\boldsymbol{\epsilon})=1$ is compact and the continuous function $f|_{M^*}$ thus assumes a minimum value. This value is $[\gamma_p(\delta)/C_p]^p$ as noticed by identifying $\rho_i=\langle i |\rho|i\rangle$.
Let $(\boldsymbol{\rho}^*,\boldsymbol{\epsilon}^*)=(\rho_1^*,\cdots,\rho_n^*,\epsilon_1^*,\cdots,\epsilon_n^*)$ be a point at which $f|_{M'}$ assumes its minimum value. If there are $k$ nonzero $p_i^*$'s, their corresponding eigenenergies cannot all be nonzero. Since if they are, we can shift all of the eigenenergies by the lowest value among them and the new point is still in $M'$ but with smaller value of $f(\boldsymbol{\rho},\boldsymbol{\epsilon})$. This means that the ground energy level must be occupied to attain a GML limit. 
Denoting the numbers of the nonzero $\epsilon_j^*$'s, which are occupied as $\rho_j^*>0$, by $l$, we therefore have $l<k$. Then we can rewrite the functions on $[0,1]^k\times[0,2\pi]^l $ as,
\begin{equation}
\begin{aligned}
&f|_{M'}(\boldsymbol{\rho},\boldsymbol{\epsilon})=\sum_{i=1}^{l}\rho_i\epsilon_i^p,\\
&g|_{M'}(\boldsymbol{\rho},\boldsymbol{\epsilon})=\left(\sum_{i=1}^{l}\rho_i\cos\epsilon_i+\sum_{i=l+1}^{k}\rho_i\right)^2+\left(\sum_{i=1}^{l}\rho_i\sin\epsilon_i\right)^2,\\
&h|_{M'}(\boldsymbol{\rho},\boldsymbol{\epsilon})=\sum_{i=1}^{k}\rho_i.
\end{aligned}
\end{equation}
We will then show that all $\epsilon_j^*$'s with  $\rho_j^*>0$ must take the same value, which means that the state is an effective two-level state.
 The gradients of the above functions at $(\boldsymbol{\rho}^*,\boldsymbol{\epsilon}^*)$ are
\begin{equation}
\begin{aligned}
&\nabla f|_{M'}(\boldsymbol{\rho}^*,\boldsymbol{\epsilon}^*)=\sum_{i=1}^{l}(\epsilon_i^*)^p\frac{\partial}{\partial \rho_i}+\sum_{i=1}^{l}p\rho_i^*(\epsilon_i^*)^{p-1}\frac{\partial}{\partial \epsilon_i},\\
&\nabla g|_{M'}(\boldsymbol{\rho}^*,\boldsymbol{\epsilon}^*)=2\sqrt{\delta}\left(\sum_{i=1}^{l} \cos(\theta^*-\epsilon_i^*)  \frac{\partial}{\partial \rho_i^*}    +\sum_{i=l+1}^{k} \cos(\theta^*)  \frac{\partial}{\partial \rho_i^*}    +\sum_{i=1}^{l} \rho_i^*\sin(\theta^*-\epsilon_i^*)  \frac{\partial}{\partial \epsilon_i}\right),\\
&\nabla h|_{M'}(\boldsymbol{\rho}^*,\boldsymbol{\epsilon}^*)=\sum_{i=1}^{k}\frac{\partial}{\partial \rho_i},
\end{aligned}
\end{equation}
where $\theta^*$ is defined by $\sum_i\rho_i^*e^{\text{i}\epsilon_i^*}=\sqrt{\delta}e^{\text{i}\theta^*}$.According to the method of Lagrange multipliers, there are real nonzero constants $\lambda$ and $\mu$ satisfying $\nabla f|_{M^*}(\boldsymbol{\rho}^*,\boldsymbol{\epsilon}^*)=\lambda \nabla g|_{M^*}(\boldsymbol{\rho}^*,\boldsymbol{\epsilon}^*)+\mu \nabla h|_{M^*}(\boldsymbol{\rho}^*,\boldsymbol{\epsilon}^*)$ and thus we have
\begin{subequations}
\begin{align}
&2\sqrt{\delta}\lambda\cos\theta+\mu=0,\label{eq:conda}\\
&\epsilon_i^p=2\lambda\sqrt{\delta}\cos(\theta-\epsilon_i)+\mu,\label{eq:condb}\\
&p\epsilon_i^{p-1}=2\lambda\sqrt{\delta}\sin(\theta-\epsilon_i).\label{eq:condc}
\end{align}
\label{eq:condabs}
\end{subequations}
Here and in the rest part of this section, we drop the symbol $^*$ for $\epsilon_i$, $\rho_i$, $\theta$ for simplicity. The equations (\ref{eq:condb}), and (\ref{eq:condc}) hold for $i=1,2,\cdots,l$, which means that all the nonzero eigenenergies occupied satisfy the same equations. We will next show that there is at most one set of solution to be reasonable. By cancelling the parameters $\lambda$, $\mu$ in the Eqs.~(\ref{eq:condabs}), we obtain the equation below,
\begin{equation}
\begin{aligned}
\epsilon_i\sin(\theta-\epsilon_i)-p\cos(\theta-\epsilon_i)+p\cos\theta=0.
\end{aligned}
\end{equation}
Define the function $F(x)$ on $[0,2\pi]$ as,
\begin{equation}
\begin{aligned}
F(x)= x\sin(\theta-x)-p\cos(\theta-x)+p\cos\theta.
\end{aligned}
\end{equation}
We are concerning the numbers of the zero points of the equation $F(x)=0$.
The derivative of $F(x)$ is
\begin{equation}
\begin{aligned}
F'(x)=\sqrt{(1-p)^2+\epsilon^2}\sin(\theta-x+\phi),
\end{aligned}
\end{equation}
where $\phi=\arctan[x/(p-1)]$ for $p<1$, and $\phi=\arctan[x/(p-1)]+\pi$ for $p>1$.
Define the function $G(x)$ on $[0,2\pi]$ as,
\begin{equation}
\begin{aligned}
G(x)=\theta-x+\arctan[x/(p-1)].
\end{aligned}
\end{equation}
The derivative of it is
\begin{equation}
\begin{aligned}
G'(x)=-1+\frac{p-1}{(p-1)^2+x^2}.
\end{aligned}
\end{equation}
Notice that $G'(x)$ is a monotone function.

We first consider the case of $p\geq2$. To count the exact numbers of the zero points, we investigate the monotonic intervals of $F(x)$ by its derivative function $F'(x)=\sqrt{(1-p)^2+x^2}\sin[G(x)+\pi]$, whose sign depends on $G(x)$. For $p\geq2$, it can be straightforwardly show that $G'(x)\in(-1,0)$ such that the range of $G(x)$ is less than $2\pi$.
Therefore, there are at most three monotonic intervals. Assume $\theta \in(0,\pi)$, we have $F'(0)=(1-p)\sin\theta<0$ and $F(2\pi)=2\pi\sin\theta>0$, which means that the function $F(x)$ will decrease, then increase, and then decrease (or there is no the third monotonic interval) in the interval $[0,2\pi]$. The one unique zero point lies in the second monotonic interval in $(0,2\pi)$. Similar considerations can be made for $\theta\in(\pi,2\pi)$ implying the one unique zero point.
As for $\theta=0$, we have $F'(0)=0$, $F''(0)=p-2\geq0$ (if $p=2$, $F''(x)=x\sin x>0$ for $x\in(0,\pi)$, hence the function will increase strictly in $(0,x_1)$ for some $x_1>0$), $F(2\pi)=0$, $F'(2\pi)<0$, which imply that there are two monotonic intervals and none zero point in $(0,2\pi)$. Similar considerations can be made for $\theta=\pi$, which completes our proof for $p\geq2$.

Then we consider it for $p<1$.
The derivative function is $F'(x)=\sqrt{(1-p)^2+x^2}\sin[G(\epsilon)]$ and we have $G'(x)<-1$. Therefore we have at least three monotonic intervals in $[0,2\pi]$. Noticing that $\arctan[x/(p-1)]\in(-\pi/2,0)$ and $G(x)$ is monotone, there are at most four monotonic intervals. Assume $\theta\in(0,\pi)$, we have $F'(0)>0$ and $F(2\pi)>0$, which yield that there are at most two zero points in the second and third monotonic intervals respectively. Assuming the two different zero points $x_1$ and $x_2$ satisfying Eqs.(\ref{eq:condabs}), we have
\begin{equation}
\begin{aligned}
\frac{x_1^{p}}{\cos(\theta-x_1)-\cos\theta}=\frac{x_2^{p}}{\cos(\theta-x_2)-\cos\theta}.
\label{eq:cond2}
\end{aligned}
\end{equation}
We will show that the equation above cannot hold. Define function $H(x)$ on $[0,2\pi]$ as,
\begin{equation}
\begin{aligned}
H(x)=\frac{x^{p}}{\cos(\theta-x)-\cos\theta}.
\end{aligned}
\end{equation}
The derivative of it is
\begin{equation}
\begin{aligned}
H'(x)=-\frac{x^{p-1}F(x)}{[\cos(\theta-x)-\cos\theta]^2}.
\end{aligned}
\end{equation}
This means that $H(x)$ is monotone in the interval $[x_1,x_2]$ (assume $x_1<x_2$) such that Eq.~(\ref{eq:cond2}) does not hold. 
This contradiction implies that there is only one solution to be possible. Similar derivations can be deduced for $\theta\in(\pi,2\pi)$.
If $\theta=0$, we have $F(0)=0$, $F'(0)=0$, $F''(0)=p-2<0$, $F(2\pi)=0$, $F'(2\pi)<0$, such that there are three monotonic intervals and one zero point in $(0,2\pi)$. Similar derivations can be deduced for $\theta=\pi$. We have thus completed our proof for $p<1$.

The remaining question is for $1<p<2$. In this case, $G'(0)>0$ and then $G'(x)$ decreases and has a zero point $x_G=\sqrt{p-1-(p-1)^2}\in[0,0.5]$. Accordingly, $G(x)$ will increase first and then decrease. The increase of $G(x)$ might bring new monotonic intervals for $F(x)$ other than the three as for the case of $p\geq2$. Since $G(x_G)-G(0)<\pi/2<\pi$, there is at most one additional monotonic interval. In conclusion, there are at most four monotonic intervals and three different zero points in $(0,2\pi)$. Assume there are three different zero points (otherwise we can prove it following the previous argument that leads to the contradiction in Eq.~(\ref{eq:cond2})), we will then prove that only one of them is possible. Assume $\theta\in[0,\pi)$, it should be satisfying $\theta+G(x_G)-G(0)>\pi$ to have an additional monotonic interval (or we are done following the arguments for $p\geq2$), such that $\theta>\pi/2$. Therefore, we have $x_G<0.5<\pi/2<\theta$. Noticing $F'(0)<0$, the first monotonic decreasing interval of $F(x)$ must lie within $[0,x_G]$ and thus $[0,\theta]$. Then we have $F'(x=\theta)=-x<0$, $F(x=\theta)=p(-1+\cos\theta)<0$, $F'(x=\theta+\pi)=x>0$, $F(x=\theta+\pi)=p(1+\cos\theta)>0$, which means that $x=\theta$ must be on the second decreasing interval, i.e. the third monotonic interval, and there is only one zero point in $[\theta,\theta+\pi]$ and two in $[0,\theta]$. According to Eq.~(\ref{eq:condc}), any pair of zero points $x_i$ and $x_j$ for $i,j=1,2,3$ should be satisfying
\begin{equation}
\begin{aligned}
\frac{x_{i}^{p-1}}{\sin(\theta-x_i)}=\frac{x_j^{p-1}}{\sin(\theta-x_j)}.
\label{eq:cond}
\end{aligned}
\end{equation}
The zero point in $[\theta,\theta+\pi]$ has $\sin(\theta-x)>0$, while the other two have $\sin(\theta-x)<0$. It is only possible to have the two different zero points in $[0,\theta]$ (or the unique one in $[\theta,\theta+\pi]$ and we are done). However, they do not satisfy the Eq.~(\ref{eq:cond2}) as discussed. Hence, there is only one possible zero point.
Similar thoughts can be deduced to $\theta\in[\pi,2\pi)$, which completes our proof for $p<1$ and all the GML bounds.

\section{Proof of the extended dual generalized Margolus-Levitin bound}
The dual GML bounds are first considered for up-bounded systems in the reference \cite{PhysRevLett.129.140403} with the orthogonality time $\tau_{p}^*=\pi\hbar/(2^{1/p}E_p^*)$, where the dual norm of the Hamiltonian is $E_p^* = \langle (\epsilon_{\text{max}}-\hat{H})^p\rangle^{1/p}$. Here we extend them for an arbitrary overlap. Defining positive definite Hamiltonian $H^*=\epsilon_{\text{max}}-H$, we are in search of the infimum,
\begin{equation}
\gamma_p(\delta) = \mathop{\text{inf}}\limits_{(H,\rho,\tau)}\frac{2^{1/p}}{\pi\hbar}\tau\langle (H^*)^p\rangle^{1/p},
\end{equation}
where the dynamical constraint is $\delta=\text{Tr}(\rho e^{-iH\tau}\rho e^{iH\tau})=\text{Tr}(\rho e^{iH^*\tau}\rho e^{-iH^*\tau})=\text{Tr}(\rho e^{-iH^*\tau}\rho e^{iH^*\tau})$. The last equality comes from the cyclicity of the trace. Therefore, this looks exactly like the problem we discussed for the GML bounds. Following similar arguments we find that the minimum exists and is taken for two-level states.
Considering the two-level Hamiltonian $H=\epsilon_1|1\rangle\langle 1|$ and an initial state $|\psi(0)\rangle = \sqrt{1-\xi^2}|0\rangle + \xi |1\rangle$, the overlap between it and the evolved state is
\begin{equation}
|\langle \psi(0)|\psi(t)\rangle| = \left(1-2\xi^2(1-\xi^2)(1-\cos \epsilon_1t)\right)^{1/2}.
\end{equation}
By substituting $E_p^*=\epsilon_1(1-\xi^2)^{1/p}$ in replacement of $\epsilon_1$ and minimizing over $z=\xi^2$, we obtain the dual GML bounds,
\begin{equation}
|\langle \psi(0)|\psi(t)\rangle| \geq O_p(\tilde{t}^*)\equiv\mathop{\text{min}}\limits_{z\in[0,1]}\left( 1-2z(1-z)(1-\cos[(2z)^{-1/p}\tilde{t}^* \pi]) \right)^{1/2},
\label{eq:gmlbound}
\end{equation}
where $\tilde{t}^* = t /\tau_p^*$ is the reduced time.

\section{The tightness of the unified bound}
In this section, we show that the unified bound composed of the GML and dual GML bounds is tight for an arbitrary two-level system under time-independent unitary evolution at any moment before the revival time $t_{\text{re}}$ when the overlap starts to increase.
The unified bound of the family of the GML bounds is
\begin{equation}
T(\delta)=\mathop{\text{max}}\limits_{p>0}\left[\frac{\pi\hbar}{2^{1/p}E_p}\gamma_{p}(\delta)\right].
\label{eq:unified_gml}
\end{equation}
The first time when the two-level system evolves to $|\langle\psi(0)|\psi(t)\rangle|=\sqrt{\delta}$ is
\begin{equation}
t(\delta)=\frac{\hbar}{\epsilon_1}\arccos \left[\frac{\delta-1+2\rho_1(1-\rho_1)}{2\rho_1(1-\rho_1)}\right].
\label{eq:2lvl}
\end{equation}
The unified GML bound is tight if and only if the two times Eq.~(\ref{eq:unified_gml}) and Eq.~(\ref{eq:2lvl}) are equal. Substituting $E_p=\epsilon_1\rho_1^{1/p}$ into Eq.~(\ref{eq:unified_gml}), it is equivalent to prove the equation below,
\begin{equation}
\mathop{\text{max}}\limits_{p>0}\left[\frac{z^{1/p}}{\rho_1^{1/p}}\arccos \left[\frac{\delta-1+2z(1-z)}{2z(1-z)}\right]\right]
=\arccos \left[\frac{\delta-1+2\rho_1(1-\rho_1)}{2\rho_1(1-\rho_1)}\right], 
\label{eq:2lvlcond}
\end{equation}
where $z$ is defined by the constraint Eq.~(\ref{eq:constr}). Assuming that the maximum is taken for $z=\rho_1$, the equation holds and hence the maximum has indeed been taken since the LHS is less than or equal to the RHS by definition. Since the constraint (\ref{eq:constr}) is satisfied only for $\rho_1\in(0,0.5)$, the unified GML bound is tight for two-level state whose ground state population is larger. 
Furthermore, we can derive exactly which GML bound is tight at a given time before 
$t_{\text{re}}\equiv\pi\hbar/\epsilon_1$ by solving the constraint. As a result, the GML bound with
\begin{equation}
    p=\frac{(1-\rho_1)(\epsilon_1t/\hbar)\sin(\epsilon_1t/\hbar)}{(1-2\rho_1)(1-\cos(\epsilon_1t/\hbar))},
\label{tightp}
\end{equation}
is tight at time $t$ for a state $\rho_1<0.5$.
Since Eq.~(\ref{tightp}) decreases with time, the GML bounds with smaller $p$ tend to limit long time evolution and otherwise the short time evolution. At time $t=0$, Eq.~(\ref{tightp}) takes its maximum value $p_0=2(1-\rho_1)/(1-2\rho_1)>2$, which is finite for $\rho_1\neq 0.5$. In these cases, GML bounds with $p>p_0>2$ do not contribute to the unified bound.
On the other hand, at time $t=t_{\text{re}}$, we have $p=0$, which means that GML bounds with arbitrary small $p$ always contribute to the unified bound.

These conclusions can be developed to the dual GML bounds by making a replacement $\rho_1\rightarrow\rho_0$.
As for $\rho_1=0.5$, the MT bound, the GML bound and dual GML bound with $p\rightarrow\infty$ are all tight and coincide with each other. This leads to an estimate to $O_p$ as $O_p(\tilde{t})<\cos(\pi\tilde{t}/2)$.
In conclusion, the unified bound of the GML and dual GML bounds is tight for an arbitrary two-level system.

In the above discussion, we found that the GML bound with $p=2$, i.e., the quadratic GML bound, is special and useful in practice. For the convenience of using this bound, we present an explicit form for it as,
\begin{equation}
\begin{aligned}
O_2(\tilde{t})=\cos\left[ \frac{\pi}{2}\left( 1 + 0.162(1-\tilde{t})+0.0743(1-\tilde{t})^2\right)\tilde{t}^{0.9521}\right],
\end{aligned}
\end{equation}
where the error is below $5\times 10^{-4}$.

\section{Parameter regimes of the QSLs}
\begin{figure}[ht]
\centering
\includegraphics[width=1.0\textwidth]{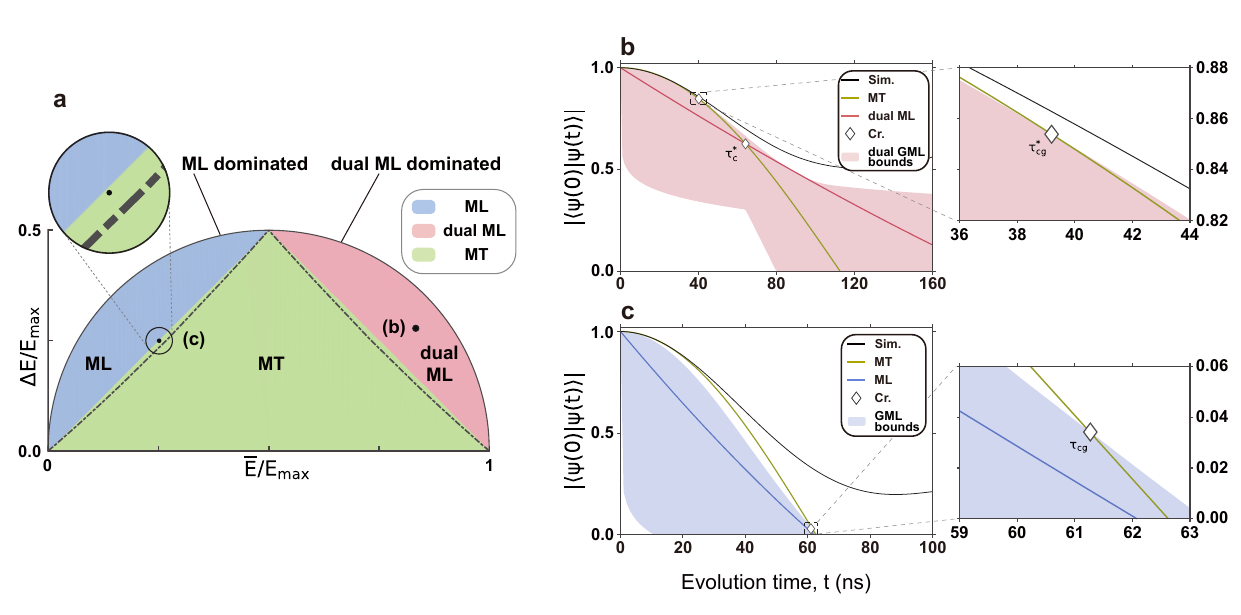}
\caption{\textbf{Parameter regimes of the QSLs.}
(a) The classification of the parameter regimes in terms of the first order moment.
We show the ML regime, the dual ML regime, and the MT regime in blue, red, and green colors, respectively.
The ML and dual ML dominated regimes are on the left and right edges of the diagram, respectively. 
The energy moments of states in (b) and (c) are marked in the diagram.
(b) QSLs in the dual ML regime. The initial state  $(\sqrt{7}|0\rangle+\sqrt{40}|1\rangle+\sqrt{115}|2\rangle)/\sqrt{162}$ is the same as it in the Fig.2(a) of the main text. 
The evolution is initially limited by the MT bound and after the generalized crossover time $\tau_{cg}^*$ (shown in the inset) limited by the unified dual GML bound. 
However, the slight modification provided by the MT bound is negligible.
(c) QSLs near the boundary of the MT regime. The initial state is $5|0\rangle+4.5|1\rangle+3.8|2\rangle+2|3\rangle+1|4\rangle$ (we neglect the normalization coefficient). 
This state is above the dashed-dot line in the MT regime as shown in the inset of (a).
The evolution is initially limited by the MT bound and later by the unified GML bound after $\tau_{cg}$ as shown in the inset.
However, the slight modification provided by the GML bounds is negligible.
}
\label{fig_regimes}
\end{figure}

In this section, we will give a more detailed discussion of the parameter regimes of the QSLs according to the energy moments, i.e., the mean and standard deviation of the energy spectra.
By assuming the ground state energy to be zero, the allowed energy moments are confined in the semicircle as shown in Fig.~\ref{fig_regimes}(a).
In the (dual) ML regime, the evolution is initially limited by the MT bound and then by the unified (dual) GML bound after the generalized crossover time $(\tau_{cg}^{*})~\tau_{cg}$, which is the time when the MT bound and the unified (dual) GML bound cross. 
We present an example in the dual ML regime in Fig.~\ref{fig_regimes}(b), where the generalized crossover is shown in the inset. 
The evolution is initially limited by MT bound and later by the dual GML bound after the generalized crossover. However, we observe that the MT bound only puts negligibly tighter limit for short time. 
On the (right) left edge of the semicircle is the (dual) ML dominated regime, which corresponds to two-level systems, where the QSL is solely governed by the unified (dual) GML bound.
In the center of the diagram is the MT regime and beneath the dashed-dot line is the MT dominated regime, where the QSL is solely governed by the MT bound. 
For states in the MT regime, however, above the dashed-dot line, there are two possibilities. 
First, the MT bound dominates the QSL, which is the normal situation in the MT regime. 
Second, the evolution is initially limited by the MT bound and later by the (dual) GML bounds. 
For this special case, we present an example in Fig.~\ref{fig_regimes}(c). 
This state is in the MT regime but above the dashed-dot line as marked in Fig.~\ref{fig_regimes}(a). 
We observe the generalized crossover $\tau_{cg}$ as shown in the inset, after which the GML bounds are tighter. Nevertheless, we observe that the unified GML bound only puts slightly tighter limit not long before the orthogonality time.
In conclusion, we can evaluate the QSLs simply by the corresponding bounds referring to the regimes.

We introduce the regimes for the coherent and squeezed states in the free resonator system with Hamiltonian $H=\hbar\nu a^\dagger a$. 
For the coherent states $|\alpha\rangle$, we have $\Bar{E}=\hbar\nu|\alpha|^2$ and $\Delta E=\hbar\nu|\alpha|$.
The critical point is met at $\tau_{\text{MT}}=\tau_{\text{ML}}$, which leads to $|\alpha|=1$. As a result, coherent states of $|\alpha|<1$ are in the ML regime, and they of $|\alpha|>1$ are in the MT regime.
For the squeezed states $|\zeta=re^{\text{i}\phi}\rangle$, we have $\Bar{E}=\hbar\nu\text{sinh}^2r$ and $\Delta E=\sqrt{2}\hbar\nu\text{sinh}r\text{cosh}r$. Since $\Delta E>\Bar{E}$ for all $r$'s, the squeezed states are always in the ML regime.

\section{device information}
The superconducting circuit is composed of a center resonator capacitively coupled to 10 transmon qubits~\cite{Wang2020}.
Each qubit owns an independent Z line for frequency modulation, a XY line for microwave excitation and a readout resonator for state discrimination. 
All readout resonators share a common transmission line for qubit state readout.

In our experiments, we select a system comprising the center resonator $R_1$ and an ancilla qubit $Q_1$ to observe the QSLs.
The ancilla qubit is used for generating the target initial states in the center resonator $R_1$ and measuring its free evolution.
The center resonator $R_1$ is a half-wavelength coplanar waveguide resonator with a fixed frequency $\omega_{R}/2\pi=5.685$ GHz.
The idle frequency $\omega_{q}/2\pi$ of the ancilla qubit $Q_1$ is set to 5.399 GHz, which is far detuned from the center resonator $R_1$ to prevent the undesired interactions.

At its idle frequency, the energy relaxation time $T_1$ and the ramsey dephasing time $T_2^{*}$ for the ancilla qubit $Q_1$ are 21.4 $\mu$s and 1.8 $\mu$s, respectively.
Meanwhile, its readout fidelity $F_0(F_1)$ is 0.9833 (0.9187) at this frequency, which is used for mitigating qubit readout errors.
For the center resonator $R_1$, the energy relaxation time $T_1$ is 11.2 $\mu$s. 
The coupling strength $\Omega/2\pi$ between the center resonator $R_1$ and the ancilla qubit $Q_1$ is 13.5 MHz, which is obtained by tuning the excited anilla qubit $Q_1$ into resonance with the center resonator $R_1$.
\begin{figure}[t]
	\includegraphics[width=1.0\textwidth]{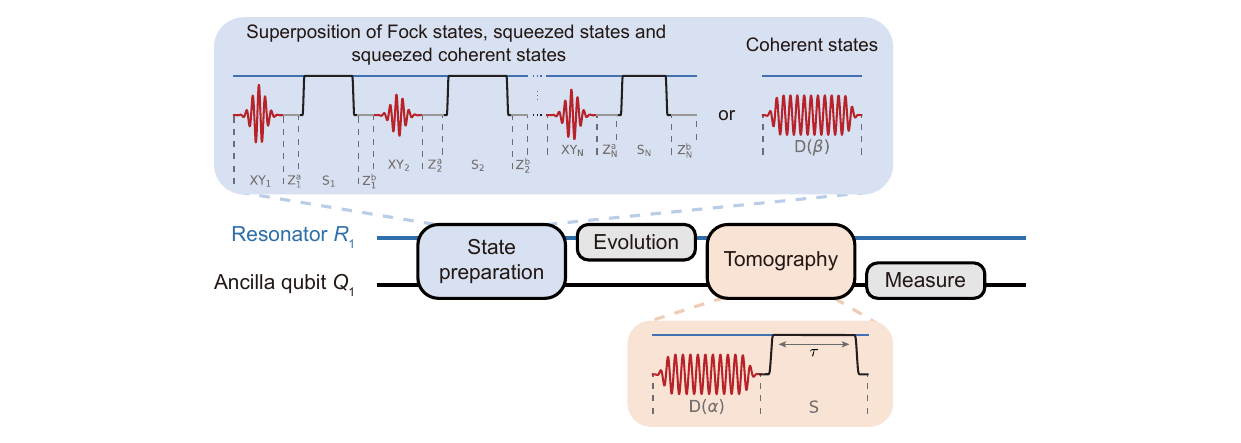}
	\caption{\label {FigS3} {\bf Sequences of the experiment.} 
    The preparation of superposition of Fock states, squeezed states and the squeezed coherent states involves a continuous process of exciting the ancilla qubit (qubit driving pulse, $\rm XY_N$) and transferring photon (qubit-resonator swap, $\rm S_N$).
    The phase rotations ($\rm Z^{a}_{N}$ and $\rm Z^{b}_{N}$) between each swap pulse $\rm S_N$ accumulate the desired phase rotations.
    The pulse sequence for generating coherent states includes only a resonant microwave pulse $ \text{D}(\rm \beta)$ with varying amplitude, executed through the XY line of the ancilla qubit $Q_1$.
    After a period of evolution, a displacement $\text{D}(\rm \alpha)$ pulse is applied, and followed by a flux bias pulse (S) that tunes the ancilla qubit $Q_1$ into resonance.
    The excitation probability $P_e(\tau)$ of the ancilla qubit $Q_1$ is obtained via a readout pulse, which is used for reconstructing the density matrix of the center resonator $R_1$.}
\end{figure}

\section{State preparation}
Preparing the arbitrary initial state can be mainly summarized as a cyclic process: first, exciting the ancilla qubit $Q_1$ into an arbitrary superposition of state $|g\rangle$ and $|e\rangle$, and then executing a controlled photon transfer between the ancilla qubit $Q_1$ and the center resonator $R_1$\cite{2009Synthesizing}. 
In our experiments, we apply a Gaussian-envelope microwave pulse to the ancilla qubit $Q_1$, yielding a $|g\rangle\rightarrow|e\rangle$ qubit transition at its idle frequency $\omega_q$.
Following the excitation of the qubit, a flux bias pulse is applied to $Q_1$, turning it into resonance with $R_1$ for a specific time to transfer the photon to the resonator.
To achieve precise single-photon transfer, we program a Gaussian-shaped overshoot at both the beginning and end of the flux bias pulse.
We note that the parameters of the excitation pulses and the flux bias pulses are carefully calibrated for different photon numbers. 

The target state of our system is given by 
\begin{equation}
|\psi\rangle=|g\rangle\otimes\sum_{n=0}^{N}c_n|n\rangle,
\end{equation}
where $c_n$ is the complex amplitude of the $n$th Fock state $|n\rangle$, and $|g\rangle$ denotes the ground state of the ancilla qubit.  
The general sequence for generating such an arbitrary target state is illustrated in Fig.\ref{FigS3}.
It involves the qubit excitation pulse ($\rm XY_N$) and the qubit-resonator photon swap ($\rm S_N$), along with two additional phase rotations,  $\rm Z^{a}_{N}$ and $\rm Z^{b}_{N}$, of the ancilla qubit, which are realized by maintaining it at the idle frequency ($\omega_q$) for a specific time.
The parameters for these experimental sequences are derived by solving the time-reversed problem\cite{2009Synthesizing}.
We optimize the phase rotations $\rm Z^{a}_{N}$ ($\rm Z^{b}_{N}$) by fine-tuning their duration times and directly probing the qubit excitation probabilities right after their following qubit-resonator photon swap (qubit excitation) pulses. The optimal duration times are determined by minimizing the difference between the measured and ideal probabilities.

In our experiment, we use the above method to generate the two- and three-level states, squeezed states, and squeezed coherent states.
Specially, the squeezed states are generated within a Hilbert space of dimension $N=6$, while the squeezed coherent states are generated with in a Hilbert space of maximum dimension $N=7$.

The preparation of coherent states is relatively simpler. 
As shown in Fig.\ref{FigS3}, a microwave pulse $\rm D(\beta)$ at the frequency of the center resonator $R_1$ is applied via the XY line of the ancilla qubit $Q_1$. 
By adjusting the amplitude of the microwave pulse $\rm D(\beta)$, the coherent states with desired photon number are generated in the center resonator $R_1$.

\section{Detection of the resonator state}
In our experiment, we detect the density matrix of the center resonator $R_1$ at different evolution times through quantum state tomography (QST), a method used for reconstructing the entire density matrix of the center resonator, utilizing the ancilla qubit $Q_1$
~\cite{LinPeng2013}.
To detect the diagonal elements of the density matrix $\rho$ of the center resonator $R_1$, the ancilla qubit $Q_1$, initially in its ground state ($|g\rangle$), is tuned in resonance with the center resonator $R_1$ for a certain time $\tau$ (see Fig. \ref{FigS3}). 
The excited state probability $P_e (\tau)$ of the ancilla qubit $Q_1$ after the interaction can be written as:
\begin{equation}
 P_{e}(\tau) = \sum_{n=0}^{N}\rho_{nn}\times P_{e}^{n}(\tau).
\label{oscillation pe}
\end{equation}
Here, $N$ is the highest occupied energy level of the resonator, $\rho_{nn}$ is the diagonal element of the density matrix of the center resonator, and $P_e^{n}(\tau) = (1/2)\left[1-\cos{(2\sqrt{n}\Omega \tau)}\right]$.
The diagonal elements are directly obtained by fitting Eq.(\ref{oscillation pe}) using the least-square method.
To reconstruct the entire density matrix of the center resonator $R_1$, we apply a displacement operator $D(\alpha)$ to the center resonator $R_1$, transforming the resonator density matrix $\rho$ into a new matrix $\rho^{\alpha}$. The new diagonal elements resulting from this transformation are given by
\begin{equation}
\rho_{nn}^{\alpha} = \langle n|D(\alpha) \rho D(\alpha)^{\dag} |n\rangle = \sum_{n_1n_2}D(\alpha)_{nn_1}D(\alpha)_{nn_2}^{*}\rho_{n_1n_2}, 
\label{transform}
\end{equation}
where $D(\alpha)=e^{\alpha a^{\dag}-\alpha^*a}$ ($\alpha=Ae^{i\varphi}$) and $D(\alpha)_{nn_1}=\langle n|D(\alpha)|n_1\rangle$. 
Applying a series of displacement $D(\alpha)$ and detecting the resulting new diagonal elements $\rho_{nn}^{\alpha}$, we can infer the entire density matrix of the center resonator $R_1$.
Experimentally, we take the energy levels $|N+1\rangle, ... ,|N^{'}\rangle$ into consideration when calculating Eq.(\ref{transform}).
We use $N^{'}=7$ for the two- and three-level states, $N^{'}=10$ for the squeezed states and squeezed coherent states, and $N^{'}=20$ for the coherent states. 
To realize the evolution of the center resonator $R_1$ under the Hamiltonian described in the main text, an additional phase $\theta = \nu t$ (where $t$ represents the evolution time) is added to the microwave pulse $\rm D(\alpha)$, with $\nu$ set to $2\pi\times4$ MHz.
\section{Supplementary data}
\begin{figure}[ht]
	\includegraphics[width=1.0\textwidth]{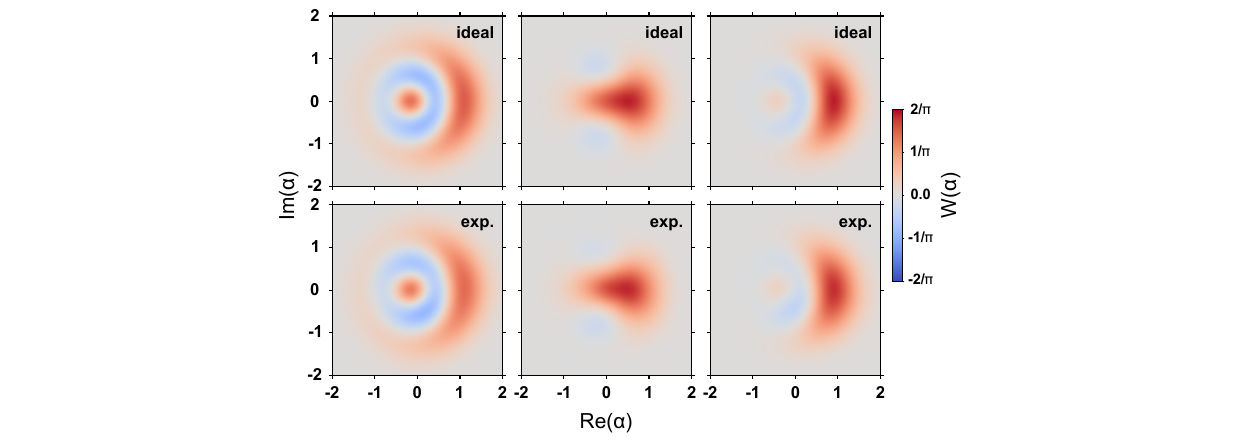}
	\caption{\label {FigS4} {\bf Wigner functions of three-level states at $t=0$ ns.} 
    The Wigner functions of three-level states, i.e. $(\sqrt{7}|0\rangle + \sqrt{40}|1\rangle + \sqrt{115}|2\rangle)/\sqrt{162}$, $(\sqrt{7}|0\rangle + |1\rangle + |2\rangle)/3$, and $(\sqrt{2}|0\rangle + \sqrt{5}|1\rangle + \sqrt{2}|2\rangle)/3$  are demonstrated from left to right. 
    The theoretical form of the Wigner functions are displayed on the top while the measured Wigner functions are shown at the bottom. 
    The fidelities between the measured density matrices and the ideal states are 0.9498$\pm$0.0015, 0.9728$\pm$0.0008, and 0.9680$\pm$0.0011, respectively.}
\end{figure}

\begin{figure}[th]
	\includegraphics[width=1.0\textwidth]{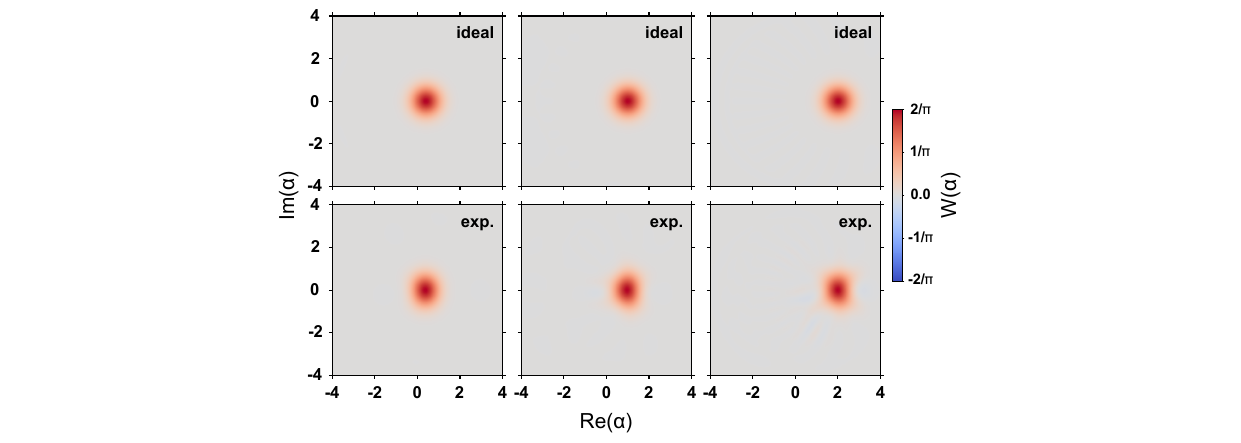}
	\caption{\label {FigS5} {\bf Wigner functions of coherent states at $t=0$ ns.} 
    Wigner functions of coherent states with $\alpha=$ 0.4, 1.0, and 2.0 are shown from left to right. 
    The fidelities between the measured density matrices and the ideal states are 0.9956$\pm$0.0001, 0.9905$\pm$0.0002, and 0.9864$\pm$0.0005 respectively.}
\end{figure}

\begin{figure}[!h]
	\includegraphics[width=1.0\textwidth]{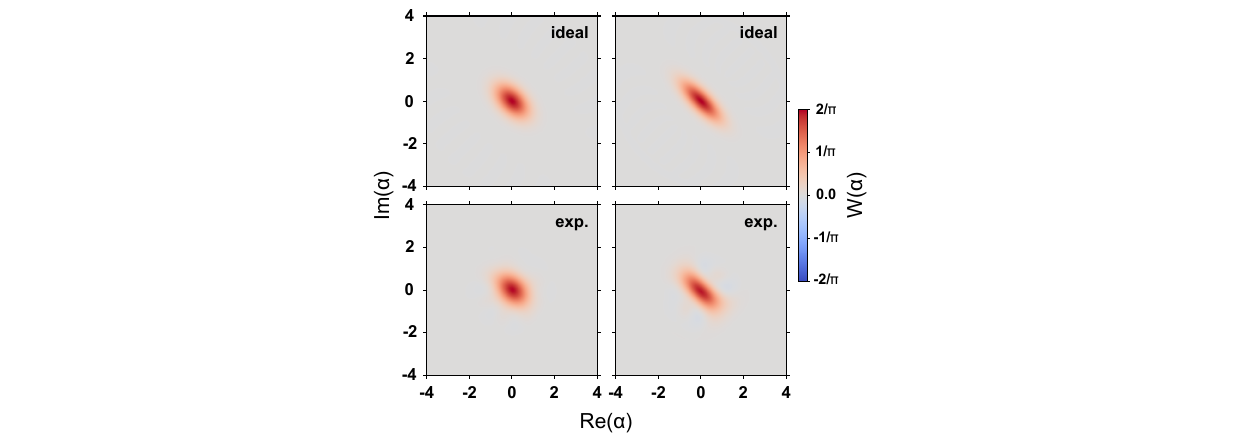}
	\caption{\label {FigS6} {\bf Wigner functions of squeezed states at $t=0$ ns.} 
    Wigner functions of squeezed states with $\zeta=0.25e^{i3\pi/2}$ and $0.5e^{i3\pi/2}$ are shown from left to right. 
    The fidelities between the measured density matrices and the ideal states are 0.9834$\pm$0.0005 and 0.9573$\pm$0.0012 respectively.}
\end{figure}

\begin{figure}[hb]
	\includegraphics[width=1.0\textwidth]{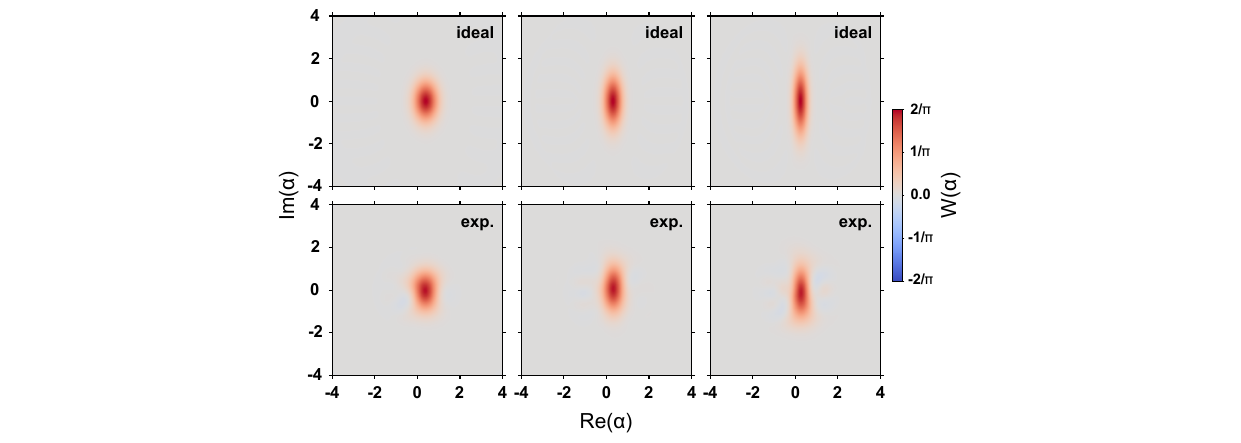}
	\caption{\label {FigS7} {\bf Wigner functions of squeezed coherent states with $\alpha=0.5$ at $t=0$ ns.} 
    Wigner functions of squeezed coherent states with the same $\alpha=0.5$ but different $\zeta=0.25, 0.5$, and $0.75$ are shown from left to right. 
    The fidelities between the measured density matrices and the ideal states are $0.9736\pm0.0006$, $0.9600\pm0.0012$, and $0.9149\pm0.0024$ respectively.}
\end{figure}

\begin{figure}[h]
	\includegraphics[width=1.0\textwidth]{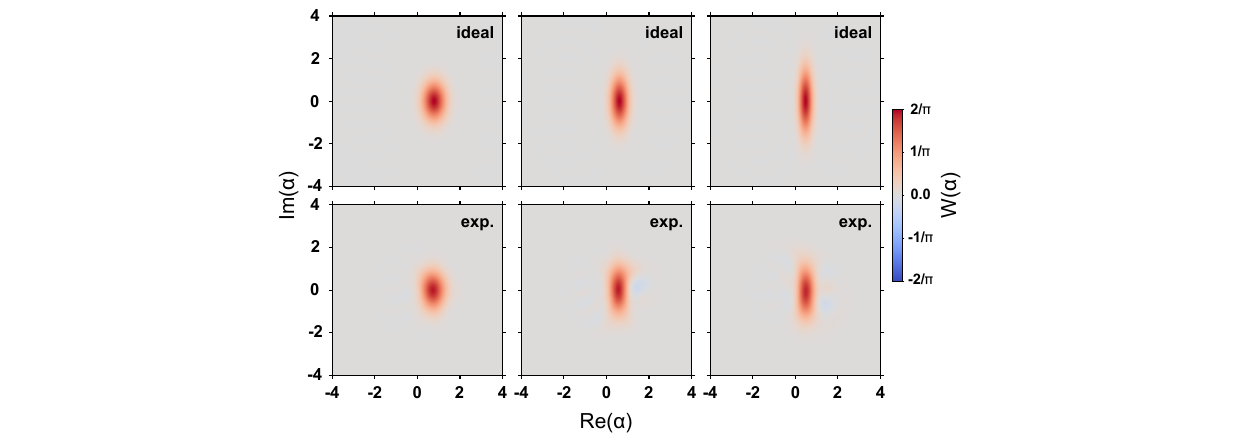}
	\caption{\label {FigS8} {\bf Wigner functions of squeezed coherent states with $\alpha=1.0$ at $t=0$ ns.} 
    Wigner functions of squeezed coherent states with the same $\alpha=1.0$ but different $\zeta=0.25, 0.5$, and $0.75$ are shown from left to right. 
    The fidelities between the measured density matrices and the ideal states are $0.9644\pm0.0009$, $0.9472\pm0.0016$, and $0.9143\pm0.0029$ respectively.}
\end{figure}

After obtaining the density matrices of the center resonator $R_1$, we can also calculate the corresponding Wigner functions $W(\alpha)=(2/\pi)\text{Tr}[D(-\alpha)\rho D(\alpha)\Pi]$ (where $\Pi=e^{i\pi a^{\dagger}a}$, is the parity operator).
The fidelities between the measured density matrices and the ideal states are calculated using $F=\langle\psi|\rho|\psi\rangle$, with ideal states expanded to the energy level $|N^{'}\rangle$.
Particularly, the fidelities of the squeezed states and squeezed coherent states are determined by comparing the measured density matrices with 20-dimensional ideal states.
The Wigner functions of the three-level states, coherent states, squeezed states and squeezed coherent states at $t=0$ ns are shown in Fig.\ref{FigS4} to \ref{FigS8}.

\subsection{Wigner functions of three-level states}
\clearpage
\subsection{Wigner functions of coherent states}

\subsection{Wigner functions of squeezed states}
\clearpage
\subsection{Wigner functions of squeezed coherent states} 

\clearpage
\subsection{Evolution of squeezed coherent states}
\begin{figure}[h]
	\includegraphics[width=1.0\textwidth]{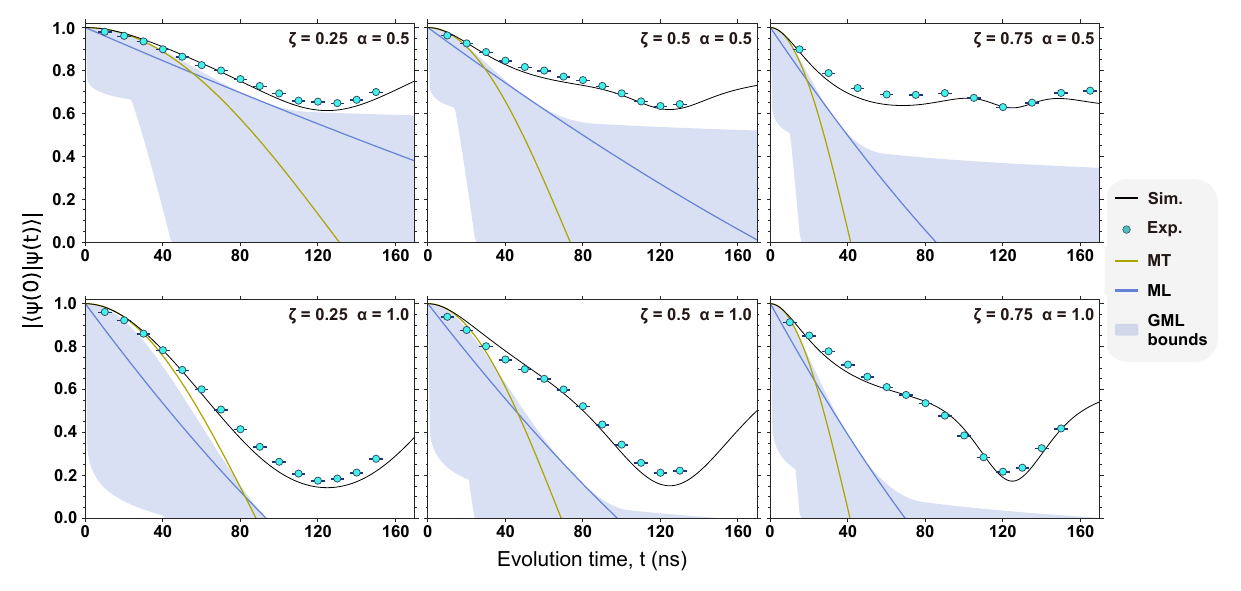}
	\caption{\label {FigS9} {\bf Evolution of squeezed coherent states $|\zeta,\alpha\rangle$ with different $\zeta$ and $\alpha$.}  The evolution of squeezed coherent states with the same $\alpha$ are shown in row, while the same $\zeta$ are displayed in column. The GML bounds are tighter for smaller $\zeta$ and $\alpha$.}
\end{figure}
\bibliography{QSLsup.bib}